\documentclass[prl,twocolumn,showpacs,aps,superscriptaddress]{revtex4}
\newcommand{\CMSG}{CoMnSi$_{1-x}$Ge$_x$}

\usepackage{amssymb,graphicx,epsf}
\usepackage{hhline,subfigure,amsmath}

\begin{document}                


\title{Negative magnetocaloric effect from highly sensitive 
metamagnetism in 
{\CMSG}}
\author{K.G. Sandeman}
\affiliation{Dept. of Materials Science and Metallurgy, University of 
Cambridge, New Museums Site, Pembroke Street, Cambridge, CB2 3QZ, UK}
\author{R. Daou}
\affiliation{Cavendish Laboratory, University of Cambridge, JJ Thompson 
Avenue, Cambridge, CB3 0HE, UK}
\author{S. \"{O}zcan}
\affiliation{Cavendish Laboratory, University of Cambridge, JJ Thompson 
Avenue, Cambridge, CB3 0HE, UK}
\author{J.H. Durrell}
\affiliation{Dept. of Materials Science and Metallurgy, University of 
Cambridge, New Museums Site, Pembroke Street, Cambridge, CB2 3QZ, UK}
\author{N.D. Mathur}
\affiliation{Dept. of Materials Science and Metallurgy, University of 
Cambridge, New Museums Site, Pembroke Street, Cambridge, CB2 3QZ, UK}
\author{D.J. Fray}
\affiliation{Dept. of Materials Science and Metallurgy, University of  
Cambridge, New Museums Site, Pembroke Street, Cambridge, CB2 3QZ, UK}
\begin{abstract}
We report a novel negative magnetocaloric effect in {\CMSG} arising 
from a metamagnetic magnetoelastic transition.  The effect is of 
relevance to magnetic refrigeration over a wide range of temperature, 
including room temperature.  In addition we report a very high shift in 
the metamagnetic transition temperature with applied magnetic field.  
This is driven by competition between antiferromagnetic and ferromagnetic 
order which can be readily tuned by applied pressure 
and compositional changes.
\end{abstract}
\maketitle

Whilst the magnetocaloric effect (MCE) has been known since
1881~\cite{warburg_1881a}, it has only recently been thought of as
providing a potential alternative to conventional gas compression
refrigeration in the room temperature range.  The conventional, positive,
MCE---where a material heats when a magnetic field is applied
adiabatically---has historically been used to achieve mK temperatures for
scientific research by demagnetisation of paramagnetic salts.  However,
the effect is largest around sharp magnetic transitions, and recent work 
has demonstrated giant MCEs near first order magnetic transitions that, by 
varying material composition and/or applied magnetic field, occur over a 
wide range of temperatures extending above room
temperature~\cite{pecharsky_1997a,tegus_2002a}.


There has already been significant progress in the design of prototype 
magnetic refrigerators~\cite{APS}, fuelled by the prediction that such 
devices could impact on carbon emissions as they are potentially 40\% more 
efficient than a conventional refrigerator~\cite{zimm_1998a}.  However, 
initial excitement arising from such developments has been tempered by two 
factors: the size
of the magnetic fields required and the cost of the magnetocaloric
refrigerants.  Ideally, permanent magnets (of strength below 2~Tesla)  
should be used.  In contrast, many prototype refrigerators have used high
fields generated by superconducting coils.  On the second point, high
purity gadolinium, on which several proposed magnetocaloric alloys are
based, has a cost of the order of \$500/kg.  Less expensive alternative
refrigerants suffer from other problems: martensitic Heusler alloys such
as Ni$_{2+x}$Mn$_{1-x}$Ga and Ni$_{2+x}$Mn$_{1-x}$Sn have a large magnetic
hysteresis~\cite{pareti_2003a,krenke_2005a};
MnFeP$_{1-x}$As$_{x}$~\cite{tegus_2002a} and MnAs-based
materials~\cite{wada_2001a} contain toxic As.  Fe$_{0.49}$Rh$_{0.51}$ is
both expensive and loses its negative MCE upon multiple cycling of the
applied field~\cite{annaorazov_1992a}.
\begin{figure}
\includegraphics[width=\columnwidth]{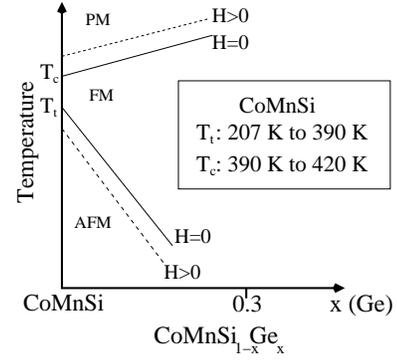}
\caption{Schematic magnetic phase diagram of {\CMSG}, after 
Nizio\l~et al.~\cite{niziol_1989a}.  Both the temperature of the 
transition between paramagnetic (PM) and ferromagnetic (FM) states, $T_c$ 
and that between the FM and antiferromagnetic (AFM) states, $T_t$ shift 
with Ge content and applied magnetic field, $H$ as shown.  Exact 
temperatures are not shown due to the variability in literature data (see 
inset). Hydrostatic pressure also reduces $T_t$ (see text).}
\label{CMSGPhaseFig}
\end{figure}

Almost all room temperature magnetocalorics exhibit a positive MCE
associated with a Curie transition.  Only the metamagnets FeRh and
Ni$_{2+x}$Mn$_{1-x}$Sn have exhibitied a significant negative
magnetocaloric effect, where the material cools when a field is applied.  
The lack of study of metamagnetic transitions by the magnetocaloric
community is surprising given that they are more likely to be first order
than their ferromagnetic cousins. In this Letter we study the
pseudoternary metamagnet {\CMSG}, a novel room temperature negative
magnetocaloric material system which addresses the issues of cost,
hysteresis and toxicity outlined above.  In particular we draw attention
to the rapid variation of its metamagnetic transition temperature, $T_t$
with magnetic field (large $|\partial T_t /\partial H|$).  This highly
desirable property usually brings about a large adiabatic temperature
change in a magnetocaloric material when it is exposed to a rapid change
in applied magnetic field over a wide range of working temperatures.  We
will show in particular that CoMnSi exhibits an MCE over a wide range of
temperatures, but this MCE is limited by such a high $|\partial T_t
/\partial H|$.  We will point to ways in which CoMnSi might be optimised
from this point of view.

The various magnetic phases of the {\CMSG} material system were examined 
by Nizio{\l} and coworkers in the 1970s and 
1980s~\cite{niziol_1979a,niziol_1989a}.  
This paper will focus on the range $x<0.1$.  CoMnSi is orthorhombic, with 
space group {\em Pnma} and exhibits
competition between helical non-collinear antiferromagnetic order and 
ferromagnetic order.  It is antiferromagnetic at low temperatures and 
shows a sample-dependent first
order metamagnetic transition to a ferromagnetic state at a transition
temperature $T_t$ of between 207~K and 360~K~\cite{niziol_1978a}.  The
ferromagnetic state has a second order $T_c$ which in much of the 
literature is at about 390~K~\cite{niziol_1989a}.  A schematic 
phase diagram of the orthorhombic phase of {\CMSG} 
for $x<0.3$ is summarised in Figure~\ref{CMSGPhaseFig}.

We concentrate here on the first order metamagnetic transition at $T_t$ in
{\CMSG}.  We note that other authors have found a wide variation in the
zero-field value of $T_t$.  Medvedeva quotes a value of 260~K in a 1~Tesla
field~\cite{medvedeva_1979a} from samples made by melting elemental Co, Mn
and a 1\% excess of Si together in a high frequency furnance under an
argon atmosphere.  Early work by Bi\'{n}czycka {\em et al.} found values
as low as 207~K in samples grown by melting elemental Co, Mn and Si,
followed by annealing at 1273~K and rapid
quenching~\cite{binczycka_1976a}.  The latter results were later
attributed to a lack of sample homogeneity, and a higher $T_t$ was
obtained by a change in growth method~\cite{niziol_1978a}. Specifically,
the change involved melting binary CoSi and elemental Mn, followed by
annealing at temperatures between 1000~K and 1200~K.  We note here that
the choice of annealing routine (hold temperature and rate of cooling) was
also observed to have an effect on the magnetic properties of CoMnSi as
early as 1973 in the work of Johnson and Frederick~\cite{johnson_1973a},
and was cited as the main cause for the sample dependent magnetic
behaviour in the work of Medvedeva~\cite{medvedeva_1979a}.

Thus, the precise magnetism of CoMnSi has been found to be extremely
sample dependent.  We suggest that this may be because the magnetism of
this material is highly sensitive to the separation of manganese atoms, on
which most of the magnetic moment is to be found~\cite{niziol_1978a}.  
Both small amounts of Ge substitution on the Si site and the application
of hydrostatic pressure have been shown to cause a rapid decrease in
$T_t$.  The rate of change of $T_t$ with pressure is very high:  
$dT_{t}/dp$ is between $-60$~K/GPa~\cite{medvedeva_1979a} and
$-100$~K/GPa~\cite{zach_1985a}.  Previous crystallographic work shows that
there is a volume contraction associated with the transition from the low
temperature antiferromagnetic state to the high temperature ferromagnetic
state~\cite{niziol_1978a}.  This would explain why the application of
hydrostatic pressure stabilises the ferromagnetic phase, reducing
$T_t$~\cite{medvedeva_1979a,zach_1985a}. Although Ge substitution expands
the lattice relative to stoichiometric CoMnSi, perhaps the reduction of
$T_t$ in that instance is driven by a change in the thermal expansion
properties of the material.  (For example, if the critical atomic
separation for a change in the exchange interaction is reached at a lower
temperature---see later.)  There is also the possibility of the observed 
variability in sample behaviour being controlled by atomic disorder, as 
yet unquantified.
	
In small fields, the metamagnetic transition at $T_t$ is probably to a
fan spin state of small net moment.  Previous literature indicates
that fields of around 2~Tesla are required to observe a transition at
$T_t$ in CoMnSi at 280~K~\cite{niziol_1979a} to a state
approaching a large magnetisation of 100 emu/g.  Here, we seek to obtain a
unified picture of the effects of substitution, pressure and magnetic
field on the tunability of the metamagnetic transition in a set of
identically fabricated samples of {\CMSG}.  The variability and possible
tunability of $T_t$ in CoMnSi makes this material an interesting candidate
magnetic refrigerant if we can readily alter the region of temperature
where the isothermal entropy change, $\Delta S$ is maximal and where 
the largest magnetocaloric effect is found.

Samples were prepared by induction melting pieces of elemental Mn
(99.99\%), Co (99.95\%), Si and Ge (both 99.9999\%) in 1 bar of argon.  
Weight losses were 0.3\% to 0.5\%.  All samples were annealed in evacuated
silica ampoules at 1223~K for 60 hours, and slowly cooled to room
temperature at a rate of 0.2~K per minute.  X-ray diffraction of powdered
samples showed only an orthorhombic ({\em Pnma}) phase.  Scanning electron
microscopy images of the materials showed a lack of significant contrast,
which, if present, would be indicative of compositional variations.  
These two observations suggest the absence of a second phase.  Rietveld
refinement of lattice parameters and atomic coordinates was also
performed. Magnetic measurements were performed in a vibrating sample 
magnetometer (maximum field 1.8~T) and a SQUID (maximum field 5~T).

\begin{figure}
\includegraphics[width=\columnwidth]{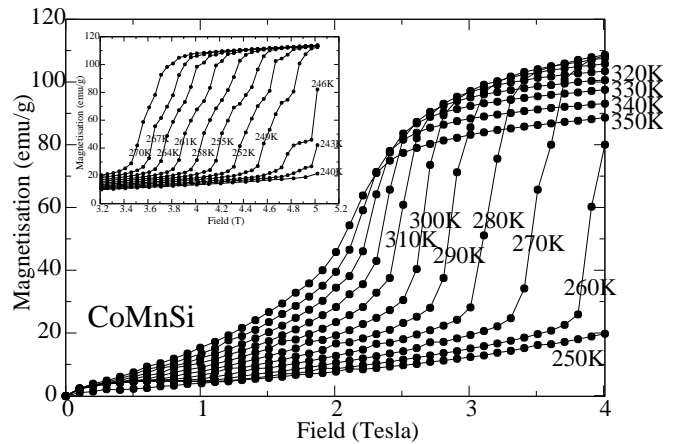}
\caption{Magnetisation versus field for CoMnSi, at 10 K~intervals in 
temperature, between 250~K and 350~K.  The inset shows the 
same measurement between 240~K and 270~K in 3~K intervals.  In this 
temperature range, the metamagnetic field is in the range 3.4~T to 5~T and 
the metamagnetic transition appears to split in two.}
\label{CoMnSiMHAllFig}
\end{figure}
\begin{figure}
\includegraphics[width=\columnwidth]{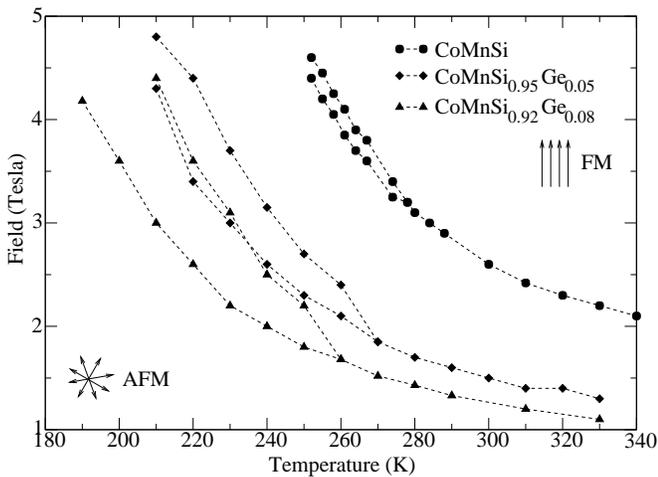}
\caption{Variation of the metamagnetic transition temperature $T_t$ with
applied field, for {\CMSG} (x=0, 0.05 and 0.08).  In each case, the 
metamagnetic transition splits into two transitions at the highest 
fields.} 
\label{MagPhaseDiag}
\end{figure}

From low field magnetisation measuremnents, we found $T_t\sim$390~K and 
$T_c\sim$420~K for CoMnSi; the highest values yet recorded for this 
material.  In Figure \ref{CoMnSiMHAllFig} we show the isothermal 
magnetisation vs
applied field for CoMnSi at temperatures between 250~K and 350~K.  Data at
each temperature was taken in increasing fields directly after zero field 
cooling from 350~K.  The first order metamagnetic transition is very 
sensitive to applied field: it shifts by 100~K in the range 2~T to 
4~T. Just as  $T_t$ is higher than hitherto measured, so the 
metamagnetic transition fields at a given temperature are larger than 
previously found. Corresponding magnetisation curves were obtained for 
{\CMSG} with $x=0.05$ or $0.08$.  These lead to the magnetic phase diagram 
shown in Fig.~\ref{MagPhaseDiag}.  In all cases, the metamagnetic field 
was taken as the point(s) of inflexion in the $M(H)$ curve.  In all 
three compounds the transition seems to split in two in the highest 
applied fields.  This is illustrated for CoMnSi in the inset to 
Figure~\ref{CoMnSiMHAllFig}, and the splitting becomes more pronounced as 
the level of Ge substitution is increased.  At this stage we cannot
establish whether the splitting of the metamagnetic transition is due to 
a lack of homogeneity or is a consequence of a high field transition to 
an intermediate, canted ferromagnetic state, as predicted for helical 
antiferromagnets by Nagamiya~\cite{nagamiya_1967a}. We note that previous 
authors associated a canted state with a much smaller magnetic crossover 
feature at lower fields~\cite{niziol_1979a}.

\begin{figure}
\includegraphics[width=\columnwidth]{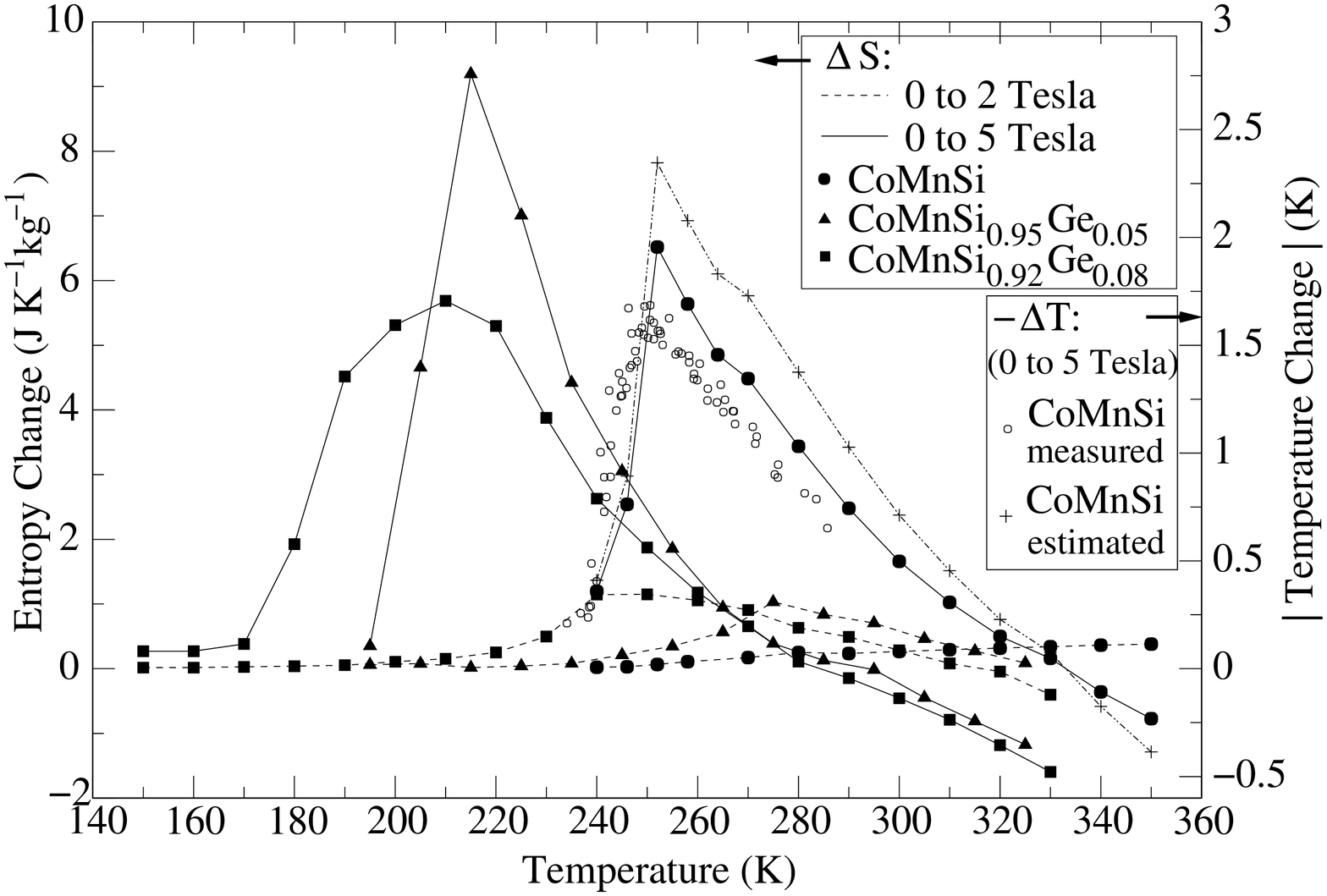}
\caption{The isothermal entropy change of {\CMSG} (x=0, 0.05 and 0.08), 
on changing the applied field from zero to either 2~Tesla or 
5~Tesla.  Also shown is the magnitude of the negative adiabatic 
temperature change, $\Delta T$, of CoMnSi when the applied field is 
raised from zero to 5~Tesla, together with an estimate of $\Delta T$ 
using our $\Delta S$ data and Eq.~\ref{DeltaT}.}
\label{CombinedDeltaSFig}
\end{figure}

We use a Maxwell relation to obtain the isothermal change in total 
entropy from the isothermal M(H) curves:
\begin{equation}
\Delta S_{\rm total}(T,\Delta H) = \int_{0}^{H_{\rm final}}{\left(\partial 
M \over \partial T\right)_{H}}\,\,dH \, . 
\label{Maxwell}
\end{equation}
This still holds true in the first order scenario if we choose to ignore 
magnetic and thermal hysteresis for the moment.  It is a fair 
approximation as the measured thermal hysteresis in CoMnSi is only 3~K at 
3~Tesla, corresponding to a small  shift in the metamagnetic transition 
field of around 0.1~Tesla.  From Equation~\ref{Maxwell}, and the M(H,T) 
data, entropy change curves for each of the three compounds were obtained 
and are shown in Figure~\ref{CombinedDeltaSFig}.
In a field change of 5~Tesla, all three compounds display a large, broad,
positive isothermal entropy change associated with $T_t$ and the onset of a 
negative change associated with $T_c$.  For a large entropy 
change associated with the metamagnetic transition to be observed, 
$T_t$ must be far removed from the Curie temperature.  This 
necessitates fields in excess of $\sim$2~Tesla.  Ge substitution 
reduces the fields and temperatures 
required for the metamagnetic transition relative to those in CoMnSi, as 
expected.   However, the transition is made less first order by 
substitution, so there is not a great increase in the size of $\Delta S$ 
for a given applied field.  We henceforth focus on CoMnSi, for which 
there exists the greatest amount of literature data with 
which to make comparisons.  The entropy change in CoMnSi is smaller than 
that in other metamagnets previously investigated (FeRh, Mn$_3$GaC) and 
this is consistent with another observation.  The rate at 
which the metamagnetic transition temperature changes with 
applied field is very large---as high as -50~K/T in low fields (right 
hand side of Figure~3), compared to -8~K/T for 
FeRh~\cite{annaorazov_1996a} and -5~K/T for 
Mn$_3$GaC~\cite{tohei_2003a}.  Such a large magnitude of $\partial 
T_t/\partial H$ (or, equivalently, small $\partial H_c/\partial T$) 
enables a wide range of working temperatures to be covered by a 
single material, although it also reduces the isothermal $\Delta S$, as 
given by the Clausius Clapeyron equation for first order magnetic phase 
transitions:
\begin{equation} 
\Delta S_{\rm total}(T,\Delta H)=-\Delta M 
\left({\partial H_c \over \partial T} \right) = -\Delta M
\left({\partial T_t \over \partial H} \right)^{-1}\, .
\label{CC} 
\end{equation} 
Here $\Delta M$ is the change in 
magnetisation at the transition, assumed to be independent of the strength 
of the applied field.  Once fields in 
excess of 2~Tesla are applied to {\CMSG}, the metamagnetic transition 
becomes more first order and $\partial T_t/\partial H$ decreases in 
magnitude, both factors leading to a much enhanced isothermal entropy 
change. Nevertheless, the extraordinarily large $\partial T_t/\partial H$ 
of CoMnSi in fields below 2~Tesla has a profound effect on 
the adiabatic temperature change, $\Delta T$:
\begin{equation} 
\Delta T(T,\Delta H\equiv H) \sim - {T\over C_{H}}\Delta S_{\rm 
total}(T,\Delta H\equiv H)
\label{DeltaT} 
\end{equation} 
where $C_{H}$ is the field-dependent heat capacity in the region of the 
magnetic transition.  We then see, by connection to Equation~\ref{CC}, 
that a very large $\partial T_t/\partial H$, as in the case of CoMnSi, 
severely reduces the adiabatic $\Delta T$.  We have also measured 
this adiabatic $\Delta T$ in a field change of zero to 5~Tesla over a 
temperature range of 230~K to 290~K, using a K-type thermocouple attached 
to a sample much larger than the dimensions of the thermocouple, 
all encased in teflon.  Fields were generated in an 8~Tesla Oxford 
Instruments cryostat.  As can be seen from the data in 
Figure~\ref{CombinedDeltaSFig}, the resulting $\Delta T(T)$ curve peaks at 
nearly 2~K at $\sim$250~K and is consistent with a crude estimate which 
may be obtained by using the 5~Tesla $\Delta S(T)$ curve and 
Eq.~\ref{DeltaT} with a fixed value of heat capacity, taken here to be 
700~J/kg, the value we obtain at $T_t$ in zero field measurements 
using a TA Instruments Q1000 DSC.

\begin{figure}
\includegraphics[width=\columnwidth]{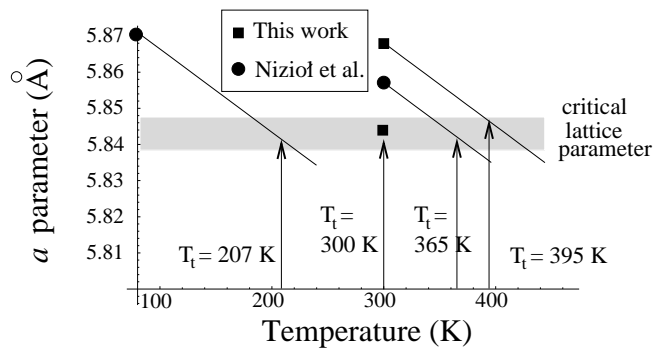}
\caption{The points show the measured $a$-axis parameter 
for samples of CoMnSi annealed in different 
ways~\cite{niziol_1979a,binczycka_1976a,niziol_1978a}.
Extrapolations of $a$ to higher temperature, based on a fixed value of 
$da/dT=-2.3\times 10^{-4}$~\AA K$^{-1}$ (extracted from Nizio\l~et 
al.~\cite{niziol_1978a}) are also shown.  The experimentally observed 
metamagnetic transition temperatures (arrows) correspond approximately to 
the same critical value of $a$ of about 5.84~\AA.} 
\label{Critical-a} 
\end{figure}

Our data, when compared with that from previous studies on CoMnSi, opens 
up the possibility of tuning the behaviour of the metamagnetic
transition in this material, perhaps through heat treatment.  We suggest
here that the fact that our material has the highest recorded zero-field
$T_t$ and the highest $H_c$ at a given temperature may be related to the
observation that it has the largest measured lattice $a$ parameter
(5.868~\AA) at room temperature.  It is known that there is a reduction in
the $a$ parameter as the temperature is increased towards the metamagnetic
transition~\cite{niziol_1978a}.  Therefore, a high room temperature value
of $a$ might yield the observed high zero-field value of $T_t$ if the
metamagnetic transition occurs at a favoured lattice spacing, as suggested
in the phenomenology of Kittel~\cite{kittel_1960a}.  This is shown
graphically in Figure \ref{Critical-a} where extrapolations of measured
lattice $a$ parameter to the metamagnetic transition temperature yield
approximately the same critical value of $a$.  
We include in this plot two of our CoMnSi samples;  one slowly cooled 
after annealing (our usual heat
treatment), and a second sample quenched instead of slowly cooled, which
had a broad metamagnetic transition at around 300~K and a reduced 
room temperature lattice $a$ parameter of 5.846~\AA.
The parameter leading to the differences between samples in the literature
and in this study may be the annealing method.  Annealing is made
necessary in this material because of a structural phase transition from a
hexagonal phase at around 1100~K encountered on cooling from the molten
state during synthesis~\cite{niziol_1989a}.  Documented annealing
temperatures vary considerably, and there is incomplete information in the
literature about the rates of cooling used.  It may be possible that
different hold temperatures and cooling rates freeze in 
different lattice strains, altering the separation of Mn atoms, 
and thereby the  sensitive metamagnetic properties of CoMnSi.

We conclude that suitable systematic control of the magnetic exchange
interactions and/or atomic order could adjust $H_c$, $T_t$ and $|{\partial
T_t /\partial H}|$ and thereby make CoMnSi a useful new negative
magnetocaloric in relatively low magnetic fields.  We have demonstrated
that a very sensitive metamagnetic transition---unlike many of those
previously studied by the magnetocaloric community---enables a broad range
of magnetocaloric working temperatures around room temperaure to be
covered by a single material, whilst harnessing the first order nature of
the transition.

We thank J. Fenstad, L.E. Hueso, A. Mischenko and S.S. Saxena for
useful discussions, K. Roberts for help with sample preparation and G.G.  
Lonzarich for use of sample synthesis facilities.  NDM and KGS acknowledge
financial support from The Royal Society.  The work was funded through UK
EPSRC Grant GR/R72235/01.

\end{document}